\documentstyle[12pt,epsfig]{article}
\begin{document}

\begin{center}
{\Large \bf Center Vortices, Nexuses, and the Georgi-Glashow Model}\\[.2in]
John M. Cornwall*\\
{\em Department of Physics and Astronomy\\
University of  California, Los Angeles\\
Los Angeles CA 90095}\\[.2in]
{\bf Abstract}
\end{center}

In a gauge theory with no Higgs fields the mechanism for confinement is by center vortices, but in theories with adjoint Higgs fields and generic symmetry breaking, such as the Georgi-Glashow model, Polyakov showed that in d=3 confinement arises via a condensate of 't Hooft-Polyakov monopoles.  We study the connection in d=3 between pure-gauge theory and the theory with adjoint Higgs by varying the Higgs VEV $v$. As one lowers $v$ from the Polyakov semiclassical regime $v\gg g$  ($g$ is the gauge coupling) toward zero, where the unbroken theory lies, one encounters effects associated with the unbroken theory at a {\em finite} value $v\simeq g$, where dynamical mass generation of a gauge-symmetric gauge-boson mass $m\simeq g^2$ takes place, in addition to the Higgs-generated non-symmetric mass $M\simeq vg$.   This dynamical mass generation is forced by the infrared instability (in both 3 and 4 dimensions) of the pure-gauge theory.  We construct solitonic configurations of the theory with both $m,M\neq 0$ which are generically closed loops consisting of nexuses  (a class of soliton recently studied for the pure-gauge theory), each paired with an antinexus, sitting like beads on a string of center vortices with vortex fields always pointing into (out of) a nexus (antinexus); the vortex magnetic fields extend a transverse distance $1/m$.  An isolated nexus with vortices is continuously deformable from the 't Hooft-Polyakov ($m=0$) monopole to the pure-gauge nexus-vortex complex ($M=0$).    In the pure-gauge $M=0$ limit the homotopy
$\Pi_2(SU(2)/U(1))=Z_2$ (or its analog for $SU(N)$) of the 't Hooft-Polyakov monopoles is no longer applicable, and is replaced by the center-vortex homotopy $\Pi_1(SU(N)/Z_N)=Z_N$ of the center vortices.\\[.1in]

\footnoterule
\noindent *E-mail address:  Cornwall@physics.ucla.edu\\
\noindent  UCLA/98/TEP/37    \mbox{} \hfill December, 1998

\newpage

\begin{center}
{\bf I. INTRODUCTION}
\end{center}

In this paper we demonstrate a smooth transition from the Georgi-Glashow model in the semiclassical limit, where confinement was argued long ago to be due to a condensate of essentially Abelian 't Hooft-Polyakov monopoles, to the center-vortex picture of confinement as proposed for the pure-gauge theory with no Higgs symmetry breaking.  The issues raised are also relevant for an understanding of claims for understanding confinement by Abelian projection.   

In the seventies several mechanisms for gauge-theory confinement were put forth.  The first continuum mechanism to be carefully worked out was Polyakov's treatment\cite{p77} of the d=3 Georgi-Glashow model.  He showed that 't Hooft-Polyakov (TP) monopoles, associated with the breaking of $SU(2)$ to $U(1)$ by an adjoint Higgs field, condensed and confined as would be expected in the much-discussed dual superconductor picture\cite{t75}.  (To avoid confusion, we note that essentially Abelian thick vortices are invoked not only in the dual superconductivity picture, but also in the center-vortex picture put forward here. These are far from being the same; in the center-vortex picture the vortices are of magnetic character, but are electric in the dual-superconductor picture.  For more modern references to the dual superconductivity hypothesis, see, {\it e.g.}, Ref. \cite{ba98}.)  

Soon thereafter the center vortex picture\cite{c79,th79,mp,no,y80} of confinement was put forth, based on the idea that a pure-gauge ({\it i.e.}, no Higgs fields to break the gauge symmetry)  theory possessed a kind of quantum soliton which was a fat object of co-dimension two (in particular, a closed string in d=3) carrying magnetic flux quantized in the center of the gauge group.  Much of this work was lattice-oriented, but Ref.\cite{c79}, working in the continuum, argued that a dynamical gauge-boson mass was generated because of infrared-instability effects, and showed that the effective lagrangian describing this mass (a gauged non-linear sigma model) had Nielsen-Olesen-like vortices.  
The mass associated with the center vortices is not associated with gauge symmetry breaking; instead, it arises\cite{c82} as a necessary element of solving the infrared-unstable Schwinger-Dyson equations of the gauge theory.  All $N^2-1$ gauge bosons of $SU(N)$ acquire the same mass $m$.  
These vortices could link with Wilson loops and for fundamental Wilson loops whose size scales were large compared to $1/m$ (hereafter, large Wilson loops) led to topological confinement in which the vortices gave rise to a   Wilson-loop phase factor of the form of an element of the center raised to a power which was a linking number:  
$\exp (2\pi i JK/N)$.   Here the integer $J$ specifies the quantized vortex flux and $K$ is the Gauss linking number of the vortex and the Wilson loop.  Averaging over these phase fluctuations then led to an area law\cite{c79}.

It has been shown by lattice-theoretic arguments\cite{t93,tk1} that in pure-gauge $SU(2)$ {\em only} center vortices can confine, by constructing lattice actions in which by adjusting parameters it is possible to retain or exclude thick center vortices and other phenomena.  Those actions with no thick center vortices are proved not to confine, for any finite lattice spacing however small.  (Thin vortices confine, but in the small-lattice-spacing limit their action is so large that they are suppressed.)

More recently, the center-vortex picture has been revived and various groups\cite{tk,fgo} have made lattice calculations comparing the area law as computed conventionally with the area law computed in various ways.  All the ways are related to, but not identical to, the continuum phase approximation which the author has used\cite{c79}.  This approximation consists, for a given gauge configuration, of replacing the true Wilson loop value by a phase factor chosen to be the element of the center nearest to the true phase factor ({\it i.e.}, for $SU(2)$ one replaces the Wilson loop by its sign).  In the continuum it is clear that the phase approximation leaves out perimeter-law terms and short-distance contributions.  These lattice calculations show that the phase approximation exactly reproduces the full area law, but their interpretation depends markedly on exactly what version of a phase approximation one uses on the lattice.  Kov\'acs and Tomboulis\cite{tk} have studied the center-vortex picture of confinement both for $SU(2)$ and for $SU(3)$, in both cases finding excellent agreement between the fundamental area law in their phase approximation and the conventional full Wilson-loop calculation.  These authors distinguish the behavior of thick vortices (those which, in the continuum, are the ones we discuss here, with a thickness of the inverse physical mass scale) and thin vortices (one lattice spacing thick) by a cooling procedure which  destroys the lattice-scale thin vortices, which cannot survive to the continuum limit.  Not only do they find that the fundamental-loop area law is exactly reproduced by their phase approximation, they find, as expected, differences at short distances.  

On the other hand, Greensite and collaborators\cite{fgo}  use different phase approximations, in some cases not fixing a gauge and in some case fixing a gauge.  Their essential $SU(2)$ phase approximation is to replace the full fundamental Wilson loop value for a given configuration by its sign.  For the fundamental-loop area law they find perfect agreement, in either case, between the full lattice calculations and their approximations.  However, when they do not fix a gauge (so-called maximal center projection), the agreement extends to short distances as well.
It has been claimed\cite{ag,og,fgo2} on the basis of an $SU(2)$ character expansion that such agreement is an inevitable consequence, given  certain very plausible behavior of higher-$J$ Wilson loops.  They then claim that fixing the gauge to the so-called maximal center gauge is, in fact, a meaningful test of the center-vortex picture.  The argument is that the gauge fixing is a global construct which can single out thick vortices, while the phase approximations without gauge fixing are infected by lattice-scale vortices.  

It would take another paper as long as this one now is to discuss these issues thoroughly.  We will make only two comments.  The first is that while it might be true that the expectation value $\langle Z \rangle$ of the sign $Z$ of the fundamental Wilson loop $W_F$ may essentially be $\langle W_F\rangle$ itself, this in itself does not answer the interesting physical questions connected with the center-vortex picture.  One such question is why $\langle Z \rangle$ yields an area law at all (aside from lattice empirics).  It could have, for instance, yielded a perimeter law.  In fact, an area law for $\langle Z \rangle $ \cite{c79,cy,corn98} comes about because center vortices have co-dimension two, that is, they are characterized by a two-dimensional density $\rho$, with the area-law coefficient (string tension) proportional to $\rho$.\footnote{Evidence for scaling behavior of an areal density for vortices on the lattice is presented in Refs. \cite{fgo3,elrt}.}  Another question, hard to address with conventional Creutz-ratio calculations of string tensions, is the actual magnitude of the higher-$J$ Wilson loops invoked in their character expansion. 
Work is underway in the continuum center-vortex picture to study such loops, but we will not discuss it here.

Second, it should be noted that the groups who argue for the trivial equality of the phase approximation and of the full fundamental Wilson loop VEV when no gauge fixing is used have been motivated in part by developments in so-called maximal Abelian projection, or MAP.  MAP is an idea introduced long ago by 't Hooft\cite{th81}.  He proposed a special way of looking at a gauge theory, by choosing a gauge in which the gauge potentials were Abelian as nearly as possible.  Points where the eigenvalues of the gauge potential had degeneracies had to be associated with monopoles, as 't Hooft showed.  The 't Hooft gauge fixing was equivalent to breaking the gauge symmetry ($SU(N)\rightarrow U(1)^{N-1}$) with an adjoint Higgs field of generic expectation value, and is relevant to our discuss of the Georgi-Glashow model.  
Recent lattice calculations\cite{sy90} are claimed to show that confinement via monopoles can indeed be seen on the lattice by projecting gauge configurations onto the Cartan subalgebra.  This projection can be done without gauge fixing, and the same groups\cite{ag,og,fgo2} who have been concerned with center-vortex gauge fixing have argued that Abelian projection without gauge fixing also does not lead to any significant test of whether 't Hooft's MAP monopoles are involved in confinement.    They have argued that the MAP-projected theory is derived from the non-Abelian theory, and not the other way around.  The present paper gives evidence for this point of view not for MAP, but for the transition from Polyakov-like confinement to center-vortex confinement in the Georgi-Glashow model.  The above-cited authors also claim that projecting onto an Abelian ensemble is not important; other projections could be used, and argue that Abelian projection is essentially trivial, that projection itself has nothing to do with the Abelian dominance claimed to be revealed by projection, and that things are very different with gauge-fixing; only with gauge-fixing can one identify the true physical objects (such as center vortices) responsible for confinement.  However, one certainly cannot base one's ultimate understanding of confinement on gauge-dependent properties.  The arguments we give here are independent of a choice of gauge.  

Once again, we forego further detailed comment on these issues, except to note that MAP
is often done with a subsidiary calculation which minimizes an action which is essentially the adjoint-Higgs gauge-boson mass term.  Clearly MAP leads to a description of a pure-gauge theory as if it were a Georgi-Glashow model.   There is one difference, however: The gauge theory quite independently of any MAP considerations generates a dynamical mass $m$.  A MAP projection may imitate the further generation of a Higgs-mechanism mass $M$, different from $m$, as well.
So MAP pictures, in general, call for consideration of the Georgi-Glashow model and its monopoles.  One must ask whether it is really some form of essentially Abelian monopole or some form of center vortex which truly underlies confinement in either a pure-gauge theory or in the Georgi-Glashow model.  This paper argues that it is the center vortex and its near relations which are essential; in this, we agree with Refs. \cite{ag,fgo2}.

 In particular, we claim that what replaces the TP monopole for finite $m$ is the combination of nexuses\cite{c79,co98,ct86} with segments of center vortices, formed into closed loops.  These closed loops lead to confinement just as pure center-vortex loops do. Fig. 1 shows a schematic model of a nexus-antinexus pair, connecting regions of center vortex with oppositely-directed fields.\footnote{We will be more specific below what we mean by oppositely-directed fields in a non-Abelian gauge theory.  In any case, it is clear what we mean when we speak of conventional photonic fields in the Georgi-Glashow model.}   In actuality the fields extend a distance of order $1/m$ transverse to the main field direction, indicated by the lines in the figure.  

\begin{figure}
\epsfig{file=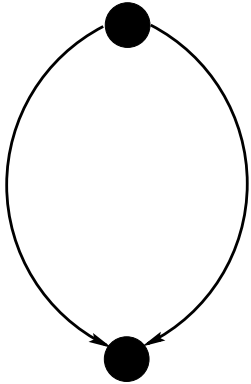,clip=,height=2in}
\caption[]{\label{cvggfig1Figure} Schematic picture of a nexus-vortex combination in $SU(2)$.
A nexus and an anti-nexus are shown as black circles.  They are joined by oppositely-directed vortex segments.}
\end{figure}

This closed loop can be interpreted as a monopole-antimonopole pair with field lines squeezed into tubes, or alternatively it can be interpreted as a center vortex with a nexus-antinexus pair (black circles) on it\cite{co98}.  Nexuses are configurations inherent to a pure-gauge theory, and we will show that they also exist in the Georgi-Glashow model, with its two different mass scales $m,M$.  In this paper we show that a nexus is the essential interpolating element between the Georgi-Glashow model in the semiclassical limit and the pure-gauge theory, where the Higgs VEV $v$ vanishes.  Generically, a nexus is a place where up to $N$ center vortices can meet, provided that their flux adds to zero (mod $N$), for gauge group $SU(N)$.
The concept of the nexus (not by that name, however) was introduced long ago\cite{c79,ct86}, but the first quantitative developments came only recently\cite{co98}.  Quite independently, Ambj\o rn and Greensite\cite{ag} have argued in favor of such configurations in the Georgi-Glashow vacuum,  and have given a cogent discussion of the differences between center vortices, the Georgi-Glashow model, and compact QED in d=3.

This picture contains, but is certainly not implied by, the picture of confinement developed by Polyakov\cite{p77} for the Georgi-Glashow model in d=3.  This model, for gauge group $SU(2)$, identifies two gauge bosons as charged, and the third as the uncharged photon, massless at the classical level.  The charged bosons pick up a mass $M=vg$ from the Higgs effect ($v$ is the Higgs VEV).  Working in the semiclassical limit where $v\gg g$, he showed that there was a condensate of TP monopoles which confined as would a dual superconductor.  He further showed that because of quantum effects the condensate density, the string tension, and an induced photon mass $m$ were all exponentially-small in the TP monpole action, which scales like $v/g\gg 1$.  

Naturally, one might expect that in the limit where the charged mass $M$ and the photon mass $m$ were the same, that is, the pure-gauge limit where the Higgs field VEV is zero and there is no symmetry breaking, that center vortices are the mechanism of confinement.  Our claim is that in comparing the d=3 Georgi-Glashow model and confinement in a pure-gauge theory, the master mechanism of confinement follows from center vortices, and that TP monopoles as they appear in the Polyakov\cite{p77}  condensate are to be understood as particular cases of the general nexus-vortex configurations we expose here.  Once the TP monopoles condense their photonic fields have a mass $m$ which is, in the semiclassical limit, small compared to the charged mass $M$.  Nevertheless, for a (fundamental) Wilson loop whose size scales are large compared to $1/m$ there is an area law of precisely the type prescribed by the center-vortex picture, following from a linkage of the nexus-vortex combination of Fig. 1 to the Wilson loop, as described above.
The flux of a single center-vortex line is half the TP monopole flux, as would be inferred from Fig. 1 by interpreting the nexus as a TP monopole.

It therefore should be possible to trace the evolution of the Georgi-Glashow model into the pure-gauge theory by varying the two masses of the theory.  There is one mass $m$ for the photon, or third component, of the gauge potential, and another mass $M$  due to symmetry breaking by a Higgs VEV.   It is useful to think of $m$ as not just a photonic mass, but as a symmetric mass present for all three gauge potentials; there is no real difference in the semiclassical regime where  $m\ll M$.  There one has the Polyakov picture, as described above.  When $M\ll m$ one has (essentially) the pure-gauge theory with its center vortices sustained by dynamical mass generation.  The two masses can be adjusted by varying the Higgs VEV $v$ from much larger than $g$ toward zero, where there is no symmetry breaking and the pure-gauge theory emerges, except for an unimportant coupling to the massless scalars.  But something perhaps unexpected arises:  Before reaching $v=0$ the Georgi-Glashow model takes on the character of the pure-gauge theory, at a critical value $v=v_c\simeq g$.  At this point infrared instability of the pure-gauge theory, in d=3,4, forces the photon mass to become of order $g^2$, the same order as the charged mass $M=vg$ becomes at the critical value.  So the Georgi-Glashow model is in the same class as the pure-gauge theory even before the symmetry breaking is restored.  In Section II we discuss this consequence of infrared instability, and give a one-loop estimate of the critical value $v_c$.  At this point there is essentially only one mass scale (even though $m\neq M$), and this scale $g^2$ gives the (inverse) distance scale for the transverse extension of the magnetic fields of Fig. 1, and there is no qualitative difference between center-vortex confinement and confinement in the Georgi-Glashow model. 

Naively it might appear that the configurations of Fig. 1 could easily be understood in some Abelian version of the theory, just as center vortices themselves and the TP monopole have a certain Abelian character.  But this is wrong; any configuration of gauge fields which has a non-zero magnetic flux over a sphere at infinity is necessarily non-Abelian, as evidenced by the homotopy $\Pi_2(SU(2)/U(1))=Z_2$.  In Section III we discuss the non-trivial constructions which lead to a qualitative description of a quantum soliton depending on the two scales $m,M$ and capable of describing a TP monopole at $m=0$ and a pure-gauge nexus-vortex combination at $M=0$.  We do not give a quantitative treatment of the soliton, which will be deferred to later work.

\begin{center}
{\bf II.  INFRARED INSTABILITY IN THE GEORGI-GLASHOW MODEL}
\end{center}

The Georgi-Glashow model is a Yang-Mills theory coupled to an adjoint Higgs field.  It can be defined for any $SU(N)$, and with generic VEVs it breaks this symmetry down to $U(1)^{N-1}$.  We will only consider it explicitly for the originally-proposed model where the gauge group $SU(2)$ is broken to $U(1)$, and we will only work out the d=3 case (as Polyakov\cite{p77} did). Two of the gauge bosons (carrying $U(1)$ charge, identified with electromagnetic (EM) charge) acquire a mass, while the third, the photon, remains massless.  This model has TP monopoles with long-range EM magnetic fields; asymptotically, the monopole fields are precisely those of the Wu-Yang singular monopoles of the corresponding pure-gauge theory.

The action for this model is:
\begin{equation}
S=\int d^3x \{\frac{1}{4g^2}(G_{ij}^a)^2 + \frac{1}{2}(D_i\psi^a)^2 + 
\frac{\lambda}{8}[(\psi^a)^2-v^2)]^2\}
\end{equation}

We will often use the conventional
antihermitean matrices
\begin{equation}
A_i=\frac{\tau^a}{2i}A_i^a,\; \psi =\frac{\tau^a}{2i}\psi^a,\;D_i
=\partial_i+A_i.
\end{equation}
The VEV of, say, $\psi^3$ is $v$; the mass $M$ of the charged gauge bosons is
$M=gv$ and the mass $M_H$ of the Higgs particle is $M_H=(\lambda )^{1/2}v$.
Mostly we are interested in the $\lambda\rightarrow \infty$ limit, where the massive Higgs particle decouples (but not the Goldstone fields).  

The semiclassical limit of the theory is $v\gg g$, or $M\gg g^2$.  The action of the TP monopole is then large:
\begin{equation}
S_{TP}=\frac{4\pi \zeta M}{g^2}=\frac{4\pi \zeta v}{g}\gg 1,
\end{equation}
where $\zeta$ is a numerical constant of order unity.  In the semiclassical theory Polyakov shows that there is a condensate of TP monopoles with a density proportional to $\exp (-S_{TP})$ which is exponentially-small in $v/g$.  The string tension is proportional to this density and therefore is exponentially-small, and the condensate causes the TP monopoles to acquire a mass $m$ which is exponentially-small too.  For all practical purposes this small mass $m$ can be ignored. 

How can one go from the Georgi-Glashow model to the pure-gauge theory, with no Higgs fields?  At the classical level, to decouple the scalar fields requires changing the sign of $v^2$ in the action (1), thereby removing the Goldstone fields which give the charged gauge bosons their mass.  If $v^2$  turns negative the symmetry is restored, and all the particles of the scalar sector acquire the same mass. Ultimately  the scalar sector can be effectively decoupled by making that mass large enough.

We will study the transition between the semiclassical Higgs regime and the pure-gauge regime by reducing $v$ toward zero from a value much larger than $g$.
Clearly, at $v=0$ the symmetry breaking is turned off, and one has a gauge theory coupled to three massless scalar fields (this is not, as we will see, an important coupling).  At first it may appear that even when $v\simeq g$ the theory looks much like the Abelian-monopole phase, with long-range EM fields for an isolated TP monopole and massive charged gauge bosons.  However, this is not so.  Because of the underlying infrared instability of the pure-gauge theory\cite{cy,co96,corn98} when the charged mass $M$ is small enough, tachyons appear in the S-matrix as calculated in one-loop\footnote{Is a one-loop result even qualitatively right?  There is some evidence that it is, from Eberlein's paper\cite{fe98} where he calculates two-loop results which are quite close to previously-calculated one-loop gap-equation results for the d=3 gauge boson mass.  However, most of these one-loop gap equation results are infected with tachyons\cite{corn98}, coming from calculated mass values which are too small to cure the infrared instability.}  perturbation theory.  The cure for these tachyons\cite{c82,chk} is a dynamically-generated mass, having nothing to do with Higgs effects, which must be large enough to overcome the tachyonic instability.  Another example of the same phenomenon occurs in Yang-Mills-Chern-Simons (YMCS) theory, where the CS term produces a mass of classical value $kg^2/4\pi$ at level $k$.  However, if $k$ is less than a critical value of order\cite{co96} 2$N$ in $SU(N)$, the tachyon persists, at least at one-loop level, and dynamical mass generation must take place.  This dynamical mass is of order $Ng^2$.  

The main technique for uncovering these results is the pinch technique (PT)\cite{c82,cp89,pa90,jw}.  In the PT a gauge-invariant gauge-boson propagator is extracted from the S-matrix by incorporating pieces of vertex, box, and other graphs which have the kinematic structure of propagator parts into the usual propagator defined by Feynman graphs. Since the S-matrix is gauge-invariant, so is the resulting propagator.  This propagator-like kinematic structure arises from pinching out certain lines in these non-propagator graphs by elementary applications of Ward identities.  

Although we will be more precise momentarily, it is useful to indicate crudely what is going on.  Roughly, the structure of the d=3 Euclidean PT propagator (denoted with a hat), when some of the gauge bosons pick up a mass $M=vg$ from the Higgs field, is (omitting inessential longitudinal terms which come from the free propagator and have no physical effect)
\begin{equation}
\hat{d}^{-1}_{ij}=(\delta_{ij}-\frac{p_ip_j}{p^2})[p^2(1-\frac{bg^2}{vg})+v^2g^2+\cdots ]
\end{equation}
where $b$ is a gauge-invariant positive number; the quantity $bg^2/vg$ is shorthand for a more complicated function as given below, but what is important for us is that it contributes negatively and that it grows as the mass $vg$ diminishes.  Evidently, if $v\leq bg$ there is a tachyonic (positive $p^2$) propagator pole.   This, then, must be removed by other effects, as indicated by the dots in (4).  Dynamical mass generation will add a mass term of order $g^4$ to the self-energy, and in combination with the $v^2g^2$ this term must self-consistently be large enough to remove the tachyon.  

Let us consider this effect for the photon in the George-Glashow model.  Even with the PT it is necessary to fix a gauge, although all gauge dependence cancels in the physically-relevant results.  We choose an $R_{\xi}$ gauge and add the gauge-fixing term $S_{GF}$ to the action (1):
\begin{equation}
S_{GF} =\int d^3x \frac{1}{2\xi}[(\partial \cdot A_2+\xi gv \psi_1)^2+
(\partial \cdot A_1-\xi gv \psi_2)^2+(\partial \cdot A_3)^2].
\end{equation}
The free charged gauge-boson propagator is
\begin{equation}
\Delta_{ij}=\frac{\delta_{ij}}{p^2+M^2} + (\xi -1)\frac{p_ip_j}{(p^2+M^2)(p^2+\xi M^2)}.
\end{equation}
As before, $M=vg$ and the Goldstone bosons and charged ghosts have squared masses $\xi (gv)^2$.
The final result for the PT one-loop inverse propagator is (again omitting free-propagator longitudinal terms):
\begin{eqnarray}
\hat{d}^{-1}_{ij}\equiv (\delta_{ij}-\frac{p_ip_j}{p^2}\hat{d}^{-1}(p^2);\\
\hat{d}^{-1}(p^2)=\{ p^2+(\frac{g^2}{4\pi})[
 (-7p+\frac{6M^2}{p})\arctan (\frac{p}{2M}) -3M]\}. \nonumber
\end{eqnarray}
Note that, as advertised, all $\xi$ dependence is gone.  
The mark of infrared instability is the $-7p$ term in (7), with its negative sign.  In the massless limit the inverse PT propagator is
\begin{equation}
\hat{d}^{-1}(p^2)=p^2-\frac{7g^2p}{8}
\end{equation}
which differs slightly from the pure-gauge PT propagator, in which -7/8 is replaced for $SU(2)$ by\cite{c82} -15/16.  The difference of +1/16 comes from the scalar fields.  The scalar contribution is infrared-stable, but it is very far in size from turning off the infrared instability of the pure-gauge theory.  

Note that the inverse PT photon propagator has a pole at $p^2=0$, as it should in perturbation theory.  But it has, for sufficiently small $v$, a tachyon as well; this tachyon also means that the zero-mass pole has negative residue.  Both phenomena are, of course, unphysical.  A quick calculation shows that this tachyon exists when
\begin{equation}
v\leq v_c;\; v_c=\frac{15g}{16\pi}.
\end{equation}

One can do a similar computation for the charged-boson PT propagator, which we do not report here; since the charged gauge bosons couple to the massive Higgs fields, the resulting condition for tachyons therefore depends on the Higgs mass.  	In any event, when $v\simeq g$, or equivalently when $M$ is small enough (of order $g^2$) tachyons appear.  These tachyons are removed in a variant of what happens in the pure-gauge theory\cite{c82,chk},  in YMCS theory\cite{co96}, or in gauge theories with Higgs symmetry breaking\cite{ch86}:  A dynamical mass of order $g^2$ is generated, so that the charged-boson mass is a combination of effects of $O(vg)$ and $O(g^2)$ and large enough to remove the tachyons.  Similarly, the photon gets a mass of $O(g^2)$.  These effects are revealed by solving dressed Schwinger-Dyson equations for the PT propagator, with the necessary full vertices approximated by a functional of the PT   propagator\cite{c82,chk,ch86,an95} which satisfies the PT Ward identities. 

We will not attempt a solution of such equations here.  Instead, we will model the results of such solutions by adding to the action (1) another mass term which is completely gauge-symmetric.  This term is just the gauged non-linear sigma model.  It has been used extensively\cite{c82,chk,an95,fe98,co96,ch86} to discuss aspects of mass generation in a pure-gauge theory, or in a Higgs model with the Higgs mass taken to infinity.

\begin{center}
{\bf  III.   FROM THE GEORGI-GLASHOW MODEL TO CENTER VORTICES}
\end{center}

As indicated in the last section, when $v$ is small enough (but not zero) there is an additional source of gauge-boson mass, coming not from Higgs effects but from the underlying infrared instability of the $SU(2)$ theory.  
We will construct an effective action, a modification of the Georgi-Glashow action (1), which represents the new source of mass.  We then search for solitons of this effective action, just as in previous works\cite{c79,co98} we sought such solitons as the center vortices themselves and the nexuses to which they can be connected.  

The first step is to construct the new effective action.  It is simply the Georgi-Glashow action (1) with an added mass term, a gauged non-linear sigma model term.  The new mass term contributes a term $m^2$ to the squared mass of all three gauge bosons, while the old Higgs mass terms contributes, as before, a mass $M^2$ to only the charged bosons.  In the limit $m=0$ we must find the original TP monopole, while in the limit $M=0$ we must find a nexus solution.  
This nexus is attached to center vortices on each side, center vortices which give a non-trivial contribution to a fundamental Wilson loop when linked to it.
As explained in Ref\cite{co98} and sketched in Fig. 1, the entire configuration consists of a nexus and an antinexus lying on a closed loop of center vortices, with the nexus and anti-nexus serving as sites for reversal of the magnetic fields of the vortices.  However, when we give explicit formulas below for solitons they will consist of a nexus at the origin, straight-line vortices, and the antinexus pushed to infinity.

\begin{center}
{\bf A.  The Two-Mass Effective Action}
\end{center}

Using the matrix notation of equation (2), we add to the action (1) a symmetric mass term:
\begin{equation}
S=\int d^3x \{\frac{-1}{2g^2}Tr(G_{ij})^2 -Tr[D_i,\psi]^2 + 
\frac{\lambda}{8}[2Tr(\psi)^2+v^2)]^2-\frac{m^2}{g^2}Tr(\tilde{A}_i)^2\}
\end{equation}
where
\begin{equation}
\tilde{A}_i=U^{-1}D_iU
\end{equation}
and $U$ is a 2$\times$2 unitary matrix representative of the group $SU(2)$.
The gauge-transformation laws for $A_i,\;U$ are:
\begin{equation}
A_i\rightarrow VA_iV^{-1}+V\partial_iV^{-1};\;U\rightarrow VU.
\end{equation}
This transformation law shows that not only is the mass term involving $m^2$ in (10) gauge-invariant, but in fact the gauge potential $\tilde{A}_i$ is locally gauge-invariant. 

The new degrees of freedom in $U$ constitute the long-range pure-gauge degrees of freedom which are responsible for confinement by linking\cite{c79}.  They are massless, and correspond to massless scalar poles which arise self-consistently in the Schwinger-Dyson equations\cite{c82,chk,an95}.  In the effective action (10) describing such solution they satisfy equations of motion which, as we show below, amount to covariant conservation for the mass sources appearing in the gauge-potential equations of motion.

We want to find solitons of effectively finite (three-dimensional) action.  One problem to be faced is the simultaneous vanishing of the two mass terms at infinite distance.  Evidently, the mass term proportional to $m^2$  vanishes when $\tilde{A}_i$ vanishes at infinity, or
\begin{equation}
x_i\rightarrow \infty:\;A_i\rightarrow U\partial_iU^{-1}.
\end{equation}
 
The long-range behavior of (14) must be compatible with the vanishing of the Higgs mass term at infinity.  It is convenient, in discussing the Higgs mass term, to introduce the modified potential $\hat{A}_i$, simply related to $\tilde{A}_i$:
\begin{equation}
\hat{A}_i=A_i+(\partial_iU)U^{-1}=U\tilde{A}_iU^{-1}.
\end{equation}
One can replace $\tilde{A}_i$ by $\hat{A}_i$ in the $m^2$ mass term of equation
(10).
It will not interfere with our main purpose to simplify the Higgs terms as much as possible, so we will in effect take the Higgs mass $M_H$ to infinity, drop the term multiplying $\lambda$ in the action (10), and replace $\Psi$ by its asymptotic value as given in (15,16) below. Given that finiteness of the $m^2$ mass term at spatial infinity requires the  requires the behavior shown in equation (13), it is standard to show that finiteness of the Higgs mass term requires:
\begin{equation}
x_i\rightarrow \infty:\;\Psi \rightarrow U\Psi_0U^{-1}
\end{equation}
where $\Psi_0$ is constant.  For example, one might choose $\Psi_0$ as: 
\begin{equation}
\Psi_0=v\frac{\tau_3}{2i}.     
\end{equation} 
Then the kinetic term for the Higgs fields, or from the point of view of the gauge bosons, the Higgs mass term in (10) becomes:
\begin{equation}
-Tr[\hat{A}_i,\Psi ]^2=-Tr[\tilde{A}_i,\Psi_0]^2.
\end{equation}

Note here the difference that the new mass term proportional to $m^2$ makes, compared to the usual TP monopole:  When $m=0$, there is no requirement that the long-range ($O(1/r)$) part of $A_i$ approach a pure gauge at spatial infinity; all that is required is that the commutator in the Higgs action in (10) vanish.  Indeed, the TP monopole in the spherical gauge, where $\Psi\rightarrow \tau_i\hat{r}_i$, is the Wu-Yang monopole, which is certainly not pure gauge.
Below we illustrate an {\it ansatz} which behaves appropriately both at $m=0$ and at $m\neq 0$.

\begin{center}
{\bf B.  From Monopole to Nexus-Vortex}
\end{center}

The action for which we seek solitonic solutions is:
\begin{equation}
S=\int d^3x \{\frac{-1}{2g^2}Tr(G_{ij})^2 -Tr[\hat{A}_i,\psi]^2  
 -\frac{m^2}{g^2}Tr(\hat{A}_i)^2\}
\end{equation}
The equations of motion for the gauge potential are:
\begin{equation}
[D_i,G_{ij}]=J_i\equiv m^2\hat{A}_i+g^2[\Psi ,[\hat{A}_i,\Psi ]].
\end{equation}
In view of the identity
\begin{equation}
[D_j,[D_i,G_{ij}]]\equiv 0
\end{equation}
it follows that
\begin{equation}
[D_i,J_i]=0;
\end{equation}
these are, as one easily shows, the equations of motion found by varying the gauge matrix $U$.  So the $U$ equations are not independent of the gauge-potential equations.

Equation (19) generically has soliton solutions, as one sees by using trial wave functions in the action (18) for $A_i,\;U$ which depend on a single spatial scale $a$.  The first term on the right-hand side of (18) scales like $1/a$ while the other two scale like $a$ times the square of some fixed mass; this action always has a minimum as $a$ is varied. 
The only questions which need investigation are whether a given trial wave function appropriately satisfies boundary conditions, and leads to no singularities in the action.\footnote{There is a short-distance singularity in $\hat{A}_i$ which gives rise to logarithmically-divergent mass contributions to the $m^2$ part of the action (18).  But since it is an essential feature of gauge-boson mass generation, whether dynamical or of Higgs nature, that the mass vanish at short distance, this singularity could be removed by a detailed consideration of the Schwinger-Dyson equations and of the effective action which they generate.  We will not do that here, nor will we be concerned further with this removable short-distance singularity.}  .  In this first investigation of the solitons we seek there is no reason to attempt stringent quantitative accuracy of the soliton action, so we will be content with finding only crude trial wave functions.  Moreover, we will spend most of our effort on insuring that the behavior of the wave function at spatial infinity is correct, since this is what determines the topology of the soliton and its confinement properties.  More detailed numerical investigations will be postponed to further work.

Our strategy is to consider solutions of the equations of motion which are determined by one or the other of the two mass terms in (18).  First we find a solution of nexus-vortex character depending only on the symmetric mass $m$.
It is simplest to display this solution in a singular gauge, where it has Dirac strings. Then we show that when $m=0$ this nexus-vortex becomes the TP monopole at distances large compared to $1/M$, or equivalently the Wu-Yang solution, in an Abelian gauge also possessing Dirac strings.  The next step is to remove the string singularities by a singular gauge transformation.  Finally, we recover a trial wave function which can be used to describe both the TP monopole and the appropriate nexus-vortex combination for generic mass values.  

Begin by recalling\cite{c79} the standard center vortex at $M$ or $\Psi=0$: 
\begin{equation}
A_i=2\pi (\frac{\tau_3}{2i})\epsilon_{ijk}\partial_j\int_{\Gamma} dz_k[
\Delta_m(x-z)-\Delta_0(x-z)]
\end{equation}
where $\Gamma$ denotes a closed contour.  In (22), $\Delta_{m,0}$ is a free scalar propagator of mass $m,0$.  The zero-mass term is a long-range pure-gauge part; it is in essence the contribution of the $U$ degrees of freedom and is responsible for confinement.  This term is the gradient of the scalar function
\begin{equation}
\Phi_{\Gamma}=\frac{\tau_3}{4i}\epsilon_{ijk}\int_Sd\sigma_{jk}\partial_k\frac{1}{|x-z|}
\end{equation}
where the integral is over a surface $S$ whose boundary is the contour $\Gamma$.
There is a $2\pi$ jump in the value of the integral in (21) for any loop which links with the contour $\Gamma$ once; this is responsible for the Dirac string in the field strength coming from this term.  

There is also a Dirac string in the $\Delta_m$ part of (22), coming from values of $x$ near the contour, where $m$ can be set to zero.  The Dirac strings from the two parts of (22) then exactly cancel, so there is no Dirac string in the field strength coming from $A_i$ as a whole. Below we will need to find another vortex which is composed of two terms, each an integral running over an open contour.  This complicates things, because as is well-known, if one tries to find a vortex for which the integral in (18) is over an open string, one finds long-range monopoles as well as  Dirac strings coming from the zero-mass propagator.  But a configuration with open strings turns out to be necessary in order to find a configuration which smoothly turns into the TP monopole at $m=0$.  It will turn out that we need a non-Abelian center vortex to do that.  The reason has to do with the flux carried by the TP monopole.

The exhibited soliton (22) is purely Abelian.  In fact, it is just a Nielsen-Olesen vortex at infinite Higgs mass.  Its Abelian nature means that it necessarily has zero flux as defined by the usual integral $\int dS_iB_i$ over a closed surface, even if the contour $\Gamma$ pierces the surface of integration.  The point is, of course, that by Stokes' theorem the flux over a {\em closed} surface must vanish, if $\vec{B}=\vec{\nabla}\times \vec{A}$.  

It follows that to make contact with the TP monopole and its non-zero flux we must find a truly non-Abelian generalization of the center vortex in (22).  Note that it is not essential that the center-vortex wave function itself be Abelian; all that matters is that the holonomy group generated by the usual loop formula
\begin{equation}
\exp (\oint dz_iA_i^{(0)}(z)),
\end{equation}
where $A_i^{(0)}$ is the long-range pure-gauge part of the center vortex configuration,
be in the center of the group (and thus Abelian).

To construct the necessary non-Abelian generalization of (22), we will abandon the long-range pure-gauge part of (22), and start with just the massive terms in Abelian form, looking for a center vortex running along the positive $z$ axis from the origin to infinity, and another running along the negative $z$ axis with oppositely-directed field strength.  The long-range pure-gauge part will be determined later.  We choose the massive part of the nexus-vortex gauge potential as:
\begin{equation}
A_i^{NV}(x)=2\pi (\frac{\tau_3}{2i})\epsilon_{ij3}\partial_j\{\int_0^{\infty} dz_3+\int_0^{-\infty}dz_3\}\Delta_m(x-z).
\end{equation}
The integrals are easily done; in cylindrical-coordinate notation one finds:
\begin{equation}
A_i^{NV}=\hat{\phi}_i(\frac{\tau_3}{2i})B;\;B\equiv \epsilon (z)[mK_1(m\rho )-J(\rho ,z;m)].
\end{equation}
Here $K_1$ is a Hankel function of imaginary argument. 
The function $J$ has the useful integral representation
\begin{equation}
J(\rho ,z;m)=\int_0^{\infty}dk\frac{k^2}{k^2+m^2}e^{-|z|( k^2+m^2)^{1/2}}J_1(k\rho ).
\end{equation} 
In spite of superficial appearances, $A_i^{NV}$ of (26,27) is a continuous differentiable function of $z$.

This (partial) vortex has Dirac strings pointing in opposite directions along the $z$ axis, because of the behavior $mK_1(m\rho )\rightarrow 1/\rho$ at short distances.  Fig. 2 sketches the general behavior of the field lines, with the singular Dirac-string fields shown as heavy lines.  This solution has zero flux, as it must because of its Abelian nature.

We first consider what happens at $m=0$.  The reason for choosing the particular form (25,26,27) is that in the massless limit it becomes the Wu-Yang monopole in a singular gauge, or equivalently the long-range part of the TP monopole:
\begin{equation}
A_i^{NV}(m=0)=\hat{\phi}_i(\frac{\tau_3}{2i\rho})\cos \theta
\end{equation}
(where $\theta$ is the polar angle).  This form of the Wu-Yang potential has two strings, just as in Fig. 2, each carrying half the Wu-Yang (or TP) magnetic flux.  The flux of each of the strings in Fig. 2 is the flux of a center vortex.

\begin{figure}
\epsfig{file=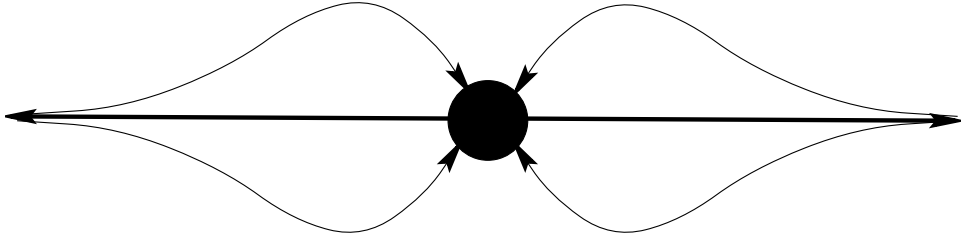,clip=,width=5in}
\caption[]{\label{cvggfig2Figure}A sketch of the field lines associated with the nexus-vortex combination of equation (25).  The heavy straight lines are Dirac-string fields.}
\end{figure}

It appears that the $m=0$ limit of the nexus-vortex combination can describe the long-range fields of a TP monopole, but at the price of introducing singular strings.  These must be gotten rid of.  We do this by finding the gauge transformations which connect the Wu-Yang or TP monopole in spherical gauge, where it has no string singularities, to the singular Abelian Wu-Yang potential (28).
The same gauge transformations occur in the deformation of the general spherical soliton to a nexus-vortex configuration\cite{co98}. 
Begin with the spherical {\it ansatz} for the TP monopole:
\begin{equation}
A_i^{TP}=\frac{\epsilon_{iak}\tau_a\hat{r}_k}{2ir}(\Phi_1(r)-1).
\end{equation}
The function $\Phi_1$ vanishes exponentially at $r=\infty$, leaving the Wu-Yang monopole to describe the long-range field.  As is well-known, this potential can be transformed to a form with an Abelian string via the gauge transformation
\begin{equation}
W=\exp (i\theta \tau\cdot\hat{\phi}/2).
\end{equation}
Applied to the TP potential (29) it yields:  
\begin{eqnarray}
A_i^{TP}\rightarrow WA_i^{TP}AW^{-1}+W\partial_iW^{-1}=\\ \nonumber
=\frac{1}{2ir}[\hat{\theta}_i\tau_2-\hat{\phi}_i\tau_1]\Phi_1+\hat{\phi}_i(\frac{\tau_3}{2i\rho})(\cos \theta -1).
\end{eqnarray}
This has a string along the negative $z$ axis, which carries the full TP monopole flux.  We can split it into two strings, each of half the flux as in Fig. 2, by the further gauge transformation $\exp (i\phi \tau_3/2)$.  
This yields:
\begin{eqnarray}
A_i^{TP}\rightarrow \frac{1}{2ir}[\hat{\theta}_i(\cos \phi \tau_2+\sin \phi \tau_1)
+\hat{\phi}_i(\sin \phi \tau_2-\cos \phi \tau_1)]\Phi_1+\\ \nonumber
+\phi_i\frac{\tau_3}{2i\rho}\cos \theta.
\end{eqnarray}

The point of this exercise is that the last term on the right-hand side of (32) is the $m=0$ limit of the nexus-vortex combination (26,27) with which we started, so we can promote that last term to finite values of $m$ easily.  Consider the trial wave function 
\begin{eqnarray}
A_i= \frac{1}{2ir}[\hat{\theta}_i(\cos \phi \tau_2+\sin \phi \tau_1)
+\hat{\phi}_i(\sin \phi \tau_2-\cos \phi \tau_1)]\Phi_1+\\ \nonumber
+\phi_i\frac{\tau_3}{2i}B,
\end{eqnarray}
differing from (32) only in the appearance of $B$ from (26,27) instead of just its massless limit.  It describes the TP monopole at $m=0$, as just discussed, and describes the nexus-vortex (26,27) at $M=0$ if we require that $\Phi$ vanish at $M=0$.  

We now wish to remove the string singularities in both the nexus-vortex and the TP monopole by applying one more singular gauge transformation $V$.  This gauge transformation will supply the long-range pure-gauge part analogous to the $\Delta_0$ term in the original center vortex (22), which cancels the string singularity of the $\Delta_m$ part.  This singular gauge transformation is not unique, but we choose:   
\begin{equation}
A_i^{NV}\rightarrow  VA_i^{NV}V^{-1}+V\partial_iV^{-1};\;V=e^{-i\phi 
\tau \cdot \hat{r}/2}.
\end{equation}
That this gauge transformation removes the strings follows simply from the observation that the string field strengths, coming from $\vec{\nabla}\times
\vec{\nabla}\phi=2\pi\hat{z}\delta (x) \delta (y)$, are multiplied by matrix coefficients to be evaluated at $x,y=0$.  We have
\begin{equation}
V\partial_iV^{-1}=\frac{\epsilon_{iak}\tau_a\hat{r}_k}{2ir}(\cos \phi -1)-
(\frac{\tau_i-\hat{r}_i\tau\cdot\hat{r}}{2ir})\sin \phi -\hat{\phi}_i
(\frac{\tau\cdot\hat{r}}{2i\rho}),
\end{equation}
where the string singularity comes from the curl of the last term on the right. 
The coefficient $\tau\cdot\hat{r}$ multiplying this singularity can be replaced by $\tau_3\cos \theta$.  Furthermore, along the $z$ axis it is clear that
$V\tau_3V^{-1}=\tau_3$.  These remarks show the cancellation of the strings in the center vortex described by (26,27).  Similarly there are no strings in the TP gauge potential, when the singular potential (32,33) is transformed by $V$ also.  The total gauge transformation going from the spherical TP potential of (29) to the final string-free potential is $V\exp (i\phi \tau_3/2)W$, but in view of the identity
\begin{equation}
e^{i\phi \tau_3/2}W=WV^{-1}
\end{equation}
the overall gauge transformation from (32) or (33) is
\begin{equation}
X\equiv VWV^{-1}.
\end{equation}

We can now suggest a non-singular trial wave function which incorporates features of both the TP monopole and the nexus-vortex combination.  It is
the potential (33) gauge-transformed by $V$ of (34).
There are no string singularities left, but there could possibly be a singularity at the origin.  Any such singularity can easily be removed by a further deformation, along the lines of Ref. \cite{co98}; we will not discuss such fine points here.  When the Higgs-generated mass $M$ vanishes one can take $\Phi_1$ to zero and recover the nexus plus vortex.  The general configuration of gauge fields is shown in Fig. 3, which is Fig. 2 with the Dirac strings missing.  Note that the holonomy of $V$ taken around a closed loop enclosing the $z$ axis is, as needed, -1, the non-trivial element of the center $Z_2$.   Because of this, the fundamental Wilson loop encircling the $z$ axis at infinite distance has the value -1, just as for the standard center-vortex picture of confinement.

\begin{figure}
\epsfig{file=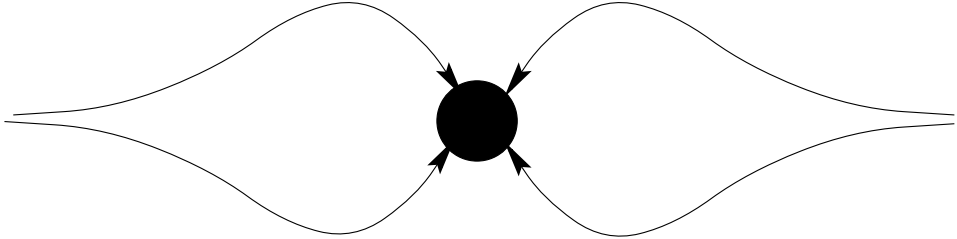,clip=,width=5in}
\caption[]{\label{cvggfig3Figure}A sketch of the field lines associated with the nexus-vortex combination of equation (33), gauge-transformed by the singular gauge transformation $V$.  The Dirac strings are gone.}
\end{figure}

The next question to ask is how the Higgs fields $\Psi$ behaves after all these gauge transformations.  Recall that in the spherical TP gauge (see (29)) the Higgs field has the kinematic structure
\begin{equation}
\Psi=v(r)(\frac{\tau\cdot\hat{r}}{2i}).
\end{equation}
The overall gauge transformation from the original spherical gauge (29) is by $X$ of (37), and the gauge-transformed Higgs kinematics follows from
\begin{equation}
X\tau\cdot\hat{r}X^{-1}=V\tau_3V^{-1}
\end{equation}
as is easily calculated.
We already know from (15) that $\Psi$ must approach a gauge transformation of a constant $\Psi_0$, with the gauge transformation precisely that which $A_i$ approaches at infinity.  This shows that in the $m\neq 0$ case the potential at infinity is:
\begin{equation}
A_i\rightarrow V\partial_iV^{-1}
\end{equation}
and therefore the matrix $U$ in the action is to be set equal to $V$, in order that both the dynamical mass term and the Higgs mass term vanish at infinity.

To summarize these arguments:
\begin{enumerate}
\item There is a nexus-vortex trial wave function  given by the gauge transform by $V=\exp (i\phi \tau\cdot\hat{r}/2)$ (see (34)) of the potential (33).  It is necessarily non-Abelian, in order to carry flux, and has non-trivial $Z_2$ holonomy.
\item  This gauge-transformed potential is free of Dirac string singularities.
\item  When the dynamical mass $m$ vanishes the result is the usual TP monopole.
\item  When the Higgs-generated mass $M$ vanishes the result is a nexus-vortex combination appropriate to the pure-gauge theory.
\item  When $m\neq 0$, the general configuration of fields is as shown in Fig. 3, with the bundle of fields on either side having the flux of a center vortex.
The transverse extent of the fields is $1/m$.
\end{enumerate}

\begin{center}
{\bf IV. CONCLUSIONS}
\end{center}

This paper has shown that the Abelian TP-monopole condensate associated with confinement in the d=3 Georgi-Glashow model is a special case of a nexus-vortex condensate, very similar to the center-vortex condensate associated with the pure-gauge theory.  Nexus-vortex combinations were first introduced for the pure-gauge theory, with equal dynamical masses for all gauge bosons; the new combinations considered here are appropriate to a broken gauge symmetry as in the Georgi-Glashow model, and have different masses $m,M$ for the neutral and charged gauge bosons, respectively.  The limit $m=0$ yields the TP monopole.  It must be remembered\cite{p77} that a condensate of such monopoles does not have strictly massless magnetic fields, although the mass $m$ induced by condensation (the Meissner effect) is exponentially small in the semiclassical regime of Polyakov's calculations.  As one moves away from the semiclassical regime of Polyakov to the quantum regime, this mass $m$ grows larger, and eventually as the Higgs symmetry breaking is turned off (by tuning the Higgs VEV $v$ toward zero) one encounters a critical value $v_c\simeq g$ at which infrared instability of the gauge theory requires generation of a dynamical mass for all three gauge bosons. This mass $m$ is of order $g^2$, and it is in addition to whatever mass $M$ (now also of order $g^2$) is induced by symmetry breaking.
For all non-zero values of $m$, including the semiclassical regime of the Georgi-Glashow model, the generic vacuum configurations responsible for confinement are of nexus-vortex type, consisting of closed loops with a nexus and anti-nexus separating vortex pieces of oppositely-directed magnetic fields; these vacuum loops have magnetic fluxes leading to a factor -1 for fundamental Wilson loops linked once to them.  The sum of the fluxes of the fields on either side of a nexus is the TP monopole flux, so the nexus introduced here is in fact the correct generalization to a vacuum condensate of the TP monopole.  For fundamental Wilson loops large compared to $1/m$ the mechanism for confinement is just that of the center-vortex picture, based on topological linking of a vacuum loop with the Wilson loop as expressed in the homotopy $\Pi_1(SU(2)/Z_2)=Z_2$.

Based on earlier work\cite{co98} on gauge-symmetric nexuses, the action of a nexus in the Georgi-Glashow model is expected to be of order $2\pi m/g^2$, multiplied by a coefficient of order unity.  In the quantum regime discussed in this paper, it is not likely that the configurational entropy of nexuses (as beads on a center-vortex string) will be big enough to overcome this action penalty, although detailed calculations need to be done to be sure. The nexus action is not necessarily large compared to entropic factors since $m$ is of order $g^2/\pi$ (again up to a coefficient of order unity).  If the action penalty is too large then the nexus and the anti-nexus on a given vacuum loop will annihilate, leaving a pure center vortex and a vacuum condensate essentially equivalent to that of the pure-gauge theory.

There are a number of interesting calculations which remain to be done, and work on some of them is in progress.  For example, when $m$ is considerably less than $M$, as in the Polyakov picture, there is a real distinction in the confinement or screening mechanisms for Wilson loops whose size is large compared to $1/M$ but small compared to $1/m$.  It is of particular interest to consider the screening mechanism for adjoint (or more generally integral-$J$) Wilson loops, where there is a real distinction between pure-gauge theory and either the Georgi-Glashow model or compact QED, as discussed in detail by Ambj\o rn and Greensite\cite{ag}.  In the adjoint Wilson loop there is only screening, not confinement\cite{co83,fgo1}, with screening produced in the Georgi-Glashow model by charged bosons of mass $M$.  As pointed out in Ref. \cite{ag}, this screening means that the Georgi-Glashow model is not really represented by a simple gas of TP monopoles.  Work is in progress to understand how the configurations displayed here impact on screening for arbitrary $SU(2)$ representations.

\vspace{.5in}

\begin{center}
{\bf ACKNOWLEDGEMENTS}
\end{center}

This work was supported in part by National Science Foundation Grant 
PHY-9531023.  I wish to thank the Aspen Center for Physics for hospitality during the summer of 1998, when part of this work was done,and to thank E. T. Tomboulis for useful discussions.

\end{document}